\documentclass[11pt,twoside]{article}
\usepackage{asp2010}

\resetcounters

\begin{document}

\allowtitlefootnote

\title{Structure of neutron stars with unified equations of state}
\author{A.~F.~Fantina$^1$, N. Chamel$^1$, J.~M. Pearson$^2$ and S. Goriely$^1$
\affil{$^1$Institut d'Astronomie et d'Astrophysique, CP-226, Universit\'e Libre de Bruxelles, 1050 Brussels, Belgium\\
$^2$D\'ept. de Physique, Universit\'e de Montr\'eal, Montr\'eal (Qu\'ebec), H3C 3J7 Canada}}

%\titlefootnote{}

\begin{abstract}
We present a set of three unified equations of states (EoSs) based on the nuclear energy-density functional (EDF) theory.
These EoSs are based on generalized Skyrme forces fitted to essentially all experimental atomic mass data and constrained 
to reproduce various properties of infinite nuclear matter as obtained from many-body calculations using realistic two- and 
three-body interactions. The structure of cold isolated neutron stars is discussed in connection with some astrophysical 
observations.
\end{abstract}

\section{Introduction}
\label{introduction}

Neutron stars, born from the catastrophic gravitational collapse of massive stars with $M\gtrsim 8 M_\odot$ at the end point 
of their evolution, are among the most compact objects in the Universe. The extreme conditions encountered in their interior make 
their description a very challenging task. Indeed, neutron stars are expected to contain very different phases of matter, from ordinary nuclei 
to homogeneous nuclear matter; their core with densities exceeding several times the nuclear matter saturation density might contain 
additional particles like hyperons or even deconfined quarks~\citep{haensel2007}. The EDF theory allows for a consistent and computationally 
tractable treatment of these various phases. We have determined the global structure of neutron stars using three different \textit{unified} 
EoSs based on the recently developed EDFs BSk19, BSk20 and BSk21~\citep{goriely2010}.

\section{Unified equations of state for cold neutron stars}
\label{eos}

The interior of a neutron star, assumed to be in full thermodynamic equilibrium at zero temperature, is composed of three main regions: 
(i) the outer crust (at densities above a few times $10^4$ g~cm$^{-3}$ and below $\sim 4 \times 10^{11}$ g~cm$^{-3}$) where nuclei arranged 
in a body-centered lattice coexist with a quantum gas of electrons, (ii) the inner crust where neutron-proton clusters coexist with both electrons and 
unbound neutrons and (iii) the liquid core (above $\sim 10^{14}$ g~cm$^{-3}$) which consists of a uniform mixture of nucleons and leptons. 
No other particles will be considered here~\citep[see][for\ a\ discussion\ of\ the\ consequence\ of\ a\ possible\ phase\ transition\ 
in\ the\ core]{chamel2012}. 

These three distinct regions are described consistently using the generalized Skyrme EDFs BSk19, BSk20 and BSk21~\citep{goriely2010}. 
These EDFs were fitted to the 2149 measured masses of nuclei with $N$ and $Z \geq 8$ given in the 2003 Atomic Mass Evaluation~\citep{audi2003},
with an rms deviation of 0.58~MeV. The masses were obtained by adding to the Hartree-Fock-Bogoliubov (HFB) energy a phenomenological Wigner 
term and correction term for the collective energy. The EDFs were also constrained to reproduce various properties of homogeneous nuclear 
matter as obtained from many-body calculations using realistic two- and three- nucleon interactions. In particular, the three different EDFs 
BSk19, BSk20 and BSk21, were fitted to three different neutron matter EoSs, reflecting the current lack of knowledge of the high-density 
behaviour of dense matter. More specifically, BSk19 (BSk21) was adjusted to the softest (stiffest) EoS of neutron matter known to us, 
whereas BSk20 was fitted to an EoS with an intermediate stiffness. %~\citep[see][for\ details]{goriely2010}.

The EoS of the outer crust has been determined in the framework of the BPS model~\citep{bps1971} using the latest experimental atomic masses 
complemented with theoretical masses obtained from our HFB mass model~\citep[see][for\ details]{pearson2011}. For the inner crust, we 
have applied the Thomas-Fermi method extended up to the 4th order with proton shell corrections added pertubatively 
via the Strutinsky integral theorem~\citep{onsi2008,pearson2012}. This method is a high-speed approximation to the self-consistent Hartree-Fock equations. 
Neutron shell corrections, which are known to be much smaller than proton shell corrections~\citep{chamel2007}, have been neglected.
In order to further reduce the computational work, nuclear clusters have been assumed to be spherical and the Wigner-Seitz approximation has
been adopted to calculate the Coulomb energy. At high densities, these clusters are found to dissolve into a plasma of neutrons, protons, 
electrons and muons thus delimiting the boundary between the inner crust and the core.

\section{Neutron star structure}
\label{ns_properties}

In this Section, we discuss the global structure of cold isolated neutron stars in connection with some astrophysical observations. For this 
purpose, we have solved the general-relativistic Tolman-Oppenheimer-Volkoff equations with our unified EoSs BSk19-20-21 using the LORENE 
library\footnote{http://www.lorene.obspm.fr}. 

\begin{figure}
\centering
\includegraphics[scale=0.4]{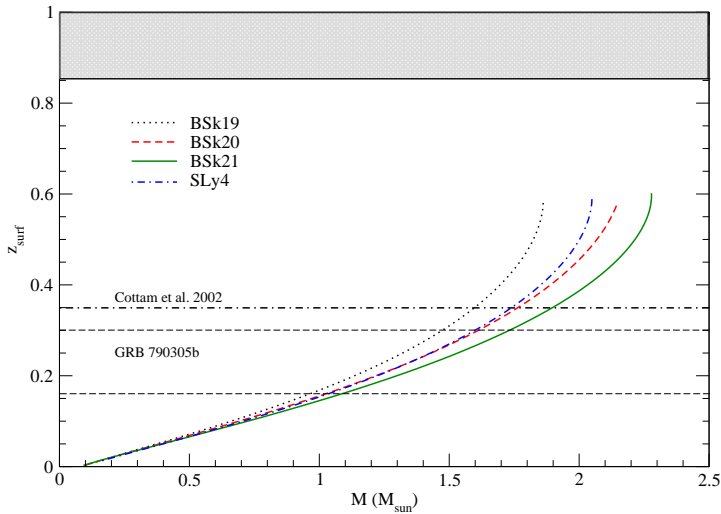}
\caption{Gravitational surface redshift as a function of the gravitational mass for the EoSs BSk19-20-21 and SLy4.}
\label{fig_zsurf}
\end{figure}

The gravitational redshift $z_{\rm surf} = (1-2GM/Rc^2)^{-1/2}-1$, where $M$ is the gravitational mass and $R$ the circumferential 
radius of the star, can be obtained from the identification of spectral features in the electromagnetic radiation from the surface 
of neutron stars. Fig.~\ref{fig_zsurf} shows the surface redshift as a function of $M$ for our three EoSs BSk19-20-21. For comparison, 
we have also plotted the redshift predicted by the SLy4 EoS~\citep{sly4}. The shaded area is prohibited by General Relativity and 
causality. The horizontal band bounded by dashed lines corresponds to the redshift estimated from the $e^+e^-$ annihilation line 
in the spectrum of the gamma-ray burst GRB 790305b~\citep{higdon1990}. Finally, the horizontal dashed-dotted line at $z_{\rm surf} = 0.35$ 
corresponds to the redshift obtained by~\citet{cottam2002} from absorption lines in the X-ray spectra of the low mass X-ray binary 
EXO 0748$-$676. The redshifts predicted by our EoSs are compatible with these observations but imply different neutron-star masses: the 
stiffer the EoS is, the larger is the mass for a given redshift. 

\begin{figure}
\centering
\includegraphics[scale=0.4]{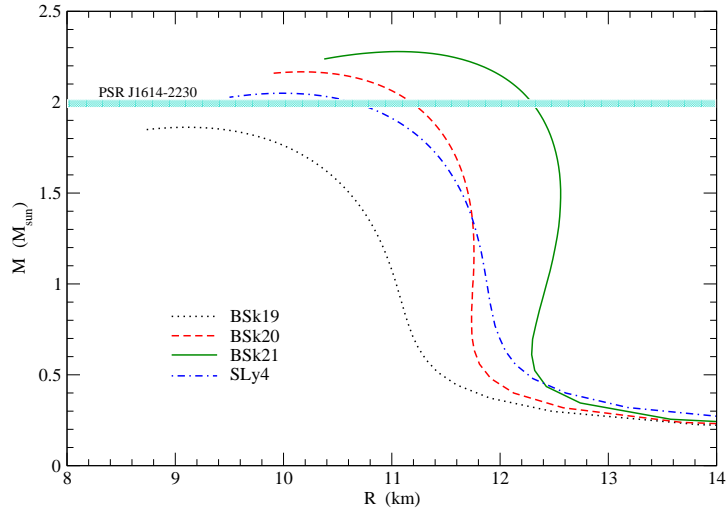}
\caption{Gravitational mass as a functional of the circumferential radius for the EoSs BSk19-20-21 and SLy4.}
\label{fig_MR}
\end{figure}

Figure~\ref{fig_MR} shows the mass-radius relation for non-rotating stars. The horizontal band around $2 M_\odot$ indicates the recently
measured value of the mass of PSR J1614$-$2230 including error bars~\citep{demorest2010}. The maximum mass of neutron stars predicted 
by our EoS BSk19 is found to be too low, even after taking into account the effects due to the pulsar rotation~\citep{chamel2011}.

\section{Conclusions}
\label{conclusion}

We have developed a set of three different unified EoSs of cold isolated neutron stars in the framework of the EDF theory. These EDFs are 
all based on generalized Skyrme forces, fitted to essentially all experimental nuclear mass data with an rms deviation falling below 
0.6 MeV~\citep{goriely2010}. At the same time, these EDFs were constrained to reproduce several properties of homogeneous nuclear matter 
(including the neutron-matter EoS) as obtained from microscopic calculations using realistic nucleon-nucleon potentials. 

These EoSs have been applied to compute the global structure of neutron stars. Even though our softest EoS BSk19 seems to be favored by 
measurements of the $\pi^-/\pi^+$ production ratio in heavy-ion collision experiments~\citep{xiao2009}, it is ruled out by the recently 
measured mass of PSR J1614$-$2230~\citep{demorest2010}. This conclusion might suggest that the dense core of neutron stars could be made 
of non-nucleonic matter~\citep{chamel2012}.

\acknowledgments This work has been supported by FNRS (Belgium), NSERC (Canada) and CompStar, 
a Research Networking Programme of the European Science Foundation.

\bibliography{fantina_talk}

\end{document}